\documentclass[pdflatex,sn-mathphys-num]{sn-jnl}
\usepackage{bookmark}

\usepackage{graphicx}%
\usepackage{multirow}%
\usepackage{amsmath,amssymb,amsfonts}%
\usepackage{amsthm}%
\usepackage[title]{appendix}%
\usepackage{xcolor}%
\usepackage{textcomp}%
\usepackage{manyfoot}%
\usepackage{booktabs}%
\usepackage{algorithm}%
\usepackage{algorithmicx}%
\usepackage{algpseudocode}%
\usepackage{listings}%

\theoremstyle{thmstyleone}%
\theoremstyle{thmstyletwo}%

\theoremstyle{thmstylethree}%

\begin{document}

\title[An Efficient Anomaly Detection Framework for Wireless Sensor Networks Using Markov Process]{An Efficient Anomaly Detection Framework for Wireless Sensor Networks Using Markov Process}


\author[1]{\fnm{Rahul} \sur{Mishra}}\email{rahulmishra@allduniv.ac.in}

\author[1]{\fnm{Sudhanshu Kumar} \sur{Jha}}\email{skjha@allduniv.ac.in}
\equalcont{These authors contributed equally to this work.}

\author[2]{\fnm{Omar Faruq} \sur{Osama}}\email{oosama@binghamton.edu  }
\equalcont{These authors contributed equally to this work.}

\author[3]{\fnm{Bishnu} \sur{Bhusal}}\email{bhusalb@missouri.com  }
\equalcont{These authors contributed equally to this work.}

\author[4]{\fnm{Sneha} \sur{Sudhakaran}}\email{ssudhakaran@fit.edu   }
\equalcont{These authors contributed equally to this work.}

\author*[5]{\fnm{Naresh} \sur{Kshetri}}\email{naresh.kshetri@rit.edu  }
\equalcont{These authors contributed equally to this work.}

\affil*[1]{\orgdiv{Department of Electronics and Communication}, \orgname{University of Allahabad}, \orgaddress{\street{Science Faculty Campus}, \city{Prayagraj}, \postcode{211002}, \state{U.P.}, \country{India}}}

\affil[2]{\orgdiv{Department of System Science $\And$ Industrial Engineering}, \orgname{Binghamton University}, \city{SUNY}, \postcode{13902},  \country{USA}}

\affil[3]{\orgdiv{Department of EE and Computer Science}, \orgname{University of Missouri}, \orgaddress{\street{Columbia}, \city{Missouri}, \postcode{65211}, \country{USA}}}

\affil[4]{\orgdiv{Department of EE and Computer Science }, \orgname{Florida Institute of Technology}, \orgaddress{\street{West Melbourne}, \city{Florida}, \postcode{32901}, \country{USA}}}

\affil[5]{\orgdiv{Department of Cybersecurity}, \orgname{Rochester Institute of Technology}, \orgaddress{\street{Rochester}, \city{New York}, \postcode{14623},  \country{USA}}}


\abstract{Wireless Sensor Networks forms the backbone of modern cyber-physical systems used in various applications such as environmental monitoring, health-care monitoring, industrial automation, and smart infrastructure. Ensuring the reliability of data collected through these networks is essential as these data may contain anomalies due to many reasons such as sensor faults, environmental disturbances, or malicious intrusions. In this paper a lightweight and interpretable anomaly detection framework based on a first-order Markov chain model has been proposed. The method discretizes continuous sensor readings into finite states and models the temporal dynamics of sensor transitions through a transition probability matrix. Anomalies are detected when observed transitions occur with probabilities below a computed threshold, allowing for real-time detection without labeled data or intensive computation. The proposed framework was validated using the Intel Berkeley Research Lab dataset, as a case study on indoor environmental monitoring demonstrates its capability to identify thermal spikes, voltage-related faults, and irregular temperature fluctuations with high precision. Comparative analysis with Z-score, Hidden Markov Model, and Autoencoder-based methods shows that the proposed Markov based framework achieves a balanced trade-off between accuracy (F1-score = 0.86), interpretability, and computational efficiency. The system’s scalability and low resource footprint highlight its suitability for large-scale and real-time anomaly detection in WSN deployments.}

\keywords{Wireless Sensor Network, Anomaly Detection, Markov Chain Model, Temporal Dynamics, Intel Berkeley Dataset}



\maketitle
\section{Introduction}\label{section-1}
\begin{figure}[h]
    \centering
    \includegraphics[width=0.6\linewidth]{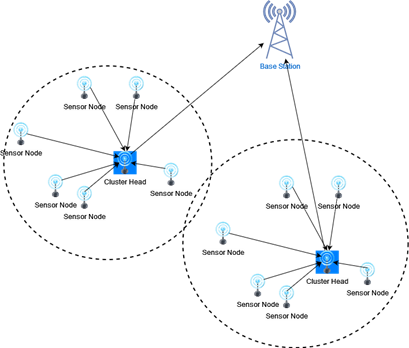}
    \caption{Hierarchical Wireless Sensor Network}
    \label{fig:fig1}
\end{figure}

Wireless Sensor Networks (WSN) are integral components in various domains, such as disaster and environmental monitoring, industrial automation, healthcare, smart cities, and many more \cite{bib1, bib2}. The WSNs are consists of distributed sensor nodes (SN) that collaborate to detect physical phenomena and report data to a central processing units called the base station (BS). The communication from SN to BS may be in a single hop, or multiple intermediate SNs are utilized to share the sensed data \cite{bib3}. Figure \ref{fig:fig1} represents a hierarchical WSN, which intricately links SN to create a sophisticated web to collect and transmit the data to the BS using intermediate cluster heads (CH) nodes or directly from the CHs. 

 Ensuring the reliability and security of WSNs is crucial, with anomaly detection playing an initial role in analyzing data streams for deviations from expected behavior \cite{bib4, bib5}. These deviations act as early warnings, signaling potential system malfunctions, unauthorized security breaches, or anomalies within wide-range applications \cite{bib6}. Sensor data anomalies can be the result of hardware malfunctions, malicious attacks, or unexpected environmental changes \cite{bib7}. Therefore, effective and lightweight anomaly detection mechanisms are essential to ensure data integrity, system reliability, and timely fault diagnosis in WSN. Traditional rule-based or thresholding methods often fail to adapt to evolving system behaviors and lack scalability for large, heterogeneous networks \cite{bib8}. Meanwhile, complex machine learning models can offer higher accuracy but typically require extensive labeled data and computational resources, which are not always feasible in real-world deployments \cite{bib9}.
 
The vigilance of anomaly detection within WSN is not merely a feature, but a critical safeguard against disruptions that could compromise the integrity and functionality of these multifaceted networks. Performing precise and instantaneous anomaly detection within WSN is a crucial work. Traditional approaches often struggle to adapt to the dynamic and complex nature of sensor data, resulting in delayed identification of emerging issues or unconventional patterns and diminishing their ability to protect the integrity and reliability of WSN \cite{bib10}. 

In this context, Markov chain-based model offers a compelling alternative by capturing patterns in sensor readings with low computational overhead and strong interpretability. A Markov chain assumes that the future state of a process depends only on the current state, not the full sequence of past states, making it particularly suitable for real-time, memory-efficient anomaly detection.

This paper presents a first-order Markov Chain-based anomaly detection framework tailored for WSNs. The approach models sensor behavior using discrete states derived from readings from the Intel Berkeley Lab dataset \cite{bib41}. The dataset is chosen as it represents a dense indoor WSN deployment for environmental monitoring, a widely used benchmark for WSN research. A transition probability matrix is constructed from historical data, and anomalies are detected by identifying unlikely or unseen state transitions. Unlike more complex models such as Hidden Markov Models (HMMs) or deep learning-based approaches, the proposed method is unsupervised, computationally efficient, and directly applicable to embedded systems. The main contribution of this paper is as follows.
\begin{itemize}
    \item A discrete state representation of sensor readings using quantile-based binning.
    \item A Markov chain model capturing normal transition dynamics in sensor behavior.
    \item An unsupervised anomaly detection algorithm based on transition probabilities.
    \item A comparative analysis against state-of-the-art techniques, highlighting the balance of simplicity, efficiency, and accuracy of the model.
\end{itemize}
The subsequent sections of this paper are organized as follows. Section 2 reviews related work, offering a summary of existing research and highlighting identified gaps. In Section 3, we explore the theoretical framework, its application in anomaly detection within WSNs, and the methodology chosen for data collection, model development, parameter estimation, and establishing the anomaly detection threshold. Section 4 details the experimental setup, presents the results obtained, and discusses the significance of the findings. Finally, in Section 5, we conclude the paper by providing insights into potential avenues for future research.

\section{Related Work} \label{section-2}

Anomaly detection in WSN has been a prominent research focus due to the increasing demand for reliable data in critical applications. The literature presents a broad spectrum of methods ranging from statistical models and machine learning algorithms to deep learning and probabilistic techniques. Through this exploration, we have not only identified key findings and methodologies from previous studies but also identified current challenges and gaps in this dynamic field. This section provides an overview of relevant anomaly detection approaches, with particular emphasis on temporal modeling, lightweight computation, and unsupervised operation.

    \subsection{Statistical and rule-based methods:} Traditional statistical methods such as mean-based, variance-based, and Mahalanobis distance-based techniques have been commonly used for anomaly detection in WSN \cite{bib15}. These methods calculate the statistical properties of sensor data and flag instances that deviate significantly from expected behavior. However, these approaches often struggle with non-Gaussian data distributions and do not capture the temporal dependencies that may exist in sensor data. The authors \cite{bib16} proposed a threshold-based method for detecting anomalies in sensor data. Although computationally efficient, they often struggle to adapt to non-stationary environments or dynamically changing network behaviors. Moreover, rule-based techniques are highly sensitive to parameter tuning and prone to high false positive rates in noisy data.
    \subsection{Machine learning-based approaches:} Machine learning methods, including Support Vector Machines (SVM), Decision Trees, and k-nearest neighbors (kNN), have been applied to WSN anomaly detection \cite{bib17,bib18}. While these techniques can handle non-linear patterns, they may require a substantial amount of labeled training data and may not effectively capture the temporal relationships in sensor data. 
    \subsection{Deep learning-based models:} Recent advancements in deep learning have introduced powerful models such as autoencoders, convolutional neural networks (CNNs) and long-short-term memory (LSTM) networks for the detection of temporal anomalies. These models can learn complex patterns from multivariate time series and capture long-range dependencies. However, their deployment in WSNs is limited due to training complexity, energy consumption, and the need for large-scale data. Interpretability remains another challenge in deep learning-based anomaly detection \cite{bib19, bib20}.
    \subsection{Probabilistic models and Markovian techniques:} Markov Chain models have gained popularity in recent years for anomaly detection in WSNs. These models offer a natural way to capture the temporal dependencies in sensor data by representing the system's state transitions over time. Researchers have explored various types of Markov Chain models, such as Hidden Markov Models (HMMs) and Markov Chain Monte Carlo (MCMC) models, to detect anomalies in WSNs \cite{bib21}. Moundounga et al. \cite{bib22} introduced a novel anomaly detection approach using HMMs in WSNs. The authors demonstrated the effectiveness of modeling sensor data as a sequence of states and using the HMM's forward-backward algorithm to detect anomalies based on state transitions. Salem \cite{bib23} proposed an HMM approach tailored for WSN, which addressed temporal dependencies in sensor data and improved detection accuracy. Kumar et al. \cite{bib24} proposed an adaptive Markov chain model that can dynamically adjust to changing network conditions and has shown improved performance in detecting anomalies under varying environmental conditions. Theerthagiri, and Prasannavenkatesan \cite{bib25} proposed a method that employs a Markov chain model, evaluating the Root Mean Square Error (RMSE) between the expected and observed values in all attributes. Recent studies \cite{bib26,bib27}, have explored hybrid approaches that combine machine learning and Markov chain models to improve anomaly detection accuracy in WSN. Their work demonstrates the benefits of leveraging the strengths of multiple techniques. Table \ref{com_tab1} provides comparison of anomaly detection approaches. 
\begin{table}[h]
\caption{Comparison of Approaches}
\label{com_tab1}
\centering
\begin{tabular}{@{}p{2cm} p{2cm} p{2cm} p{2.5cm} p{2.5cm}@{}}
\hline
\textbf{Approach} & \textbf{Model} & \textbf{Data} & \textbf{Strengths} & \textbf{Weaknesses} \\
\hline
Threshold-based & Rule-based & Any & Simple, fast & High false positives \\
SVM / One-Class SVM & Machine Learning & Labeled / Unlabeled sensor data & Good accuracy with labeled data & Sensitive to outliers, poor interpretability \\
Isolation Forest & Ensemble & Multivariate sensor data & Works without labels & Ignores temporal correlations \\
Autoencoder & Deep Learning & Historical sensor data & Learns complex nonlinear patterns & Needs large data, less interpretable \\

\hline
\end{tabular}
\end{table}

Table [2] provides information on key research endeavors in anomaly detection within WSN, using the Markov Chain Model.  Although these studies have made significant contributions to the field, several gaps and challenges remain unaddressed. In particular, a gap exists in terms of addressing the scalability and temporal dependency of existing anomaly detection techniques in WSNs, particularly when considering the dynamic and resource-constrained nature of the networks.  

\begin{table}[ht]
\caption{Summary of Anomaly Detection Studies Using Markov Chain}
\label{tab:anomaly-detection}
\centering
\begin{tabular}{@{}p{2.5cm} p{2.7cm} p{6cm}@{}}
\hline
\textbf{Objective} & \textbf{Methodology} & \textbf{Key Findings} \\
\hline
Detect anomalies in environmental monitoring & Applied Hidden Markov Models (HMMs) \cite{bib28} & Achieved 90\% accuracy in identifying abnormal environmental patterns \\
Anomaly detection for industrial automation & Extended Markov Chain to accommodate dynamic features \cite{bib29} & Successfully identified and isolated anomalous behaviours in industrial processes \\
Healthcare application: Patient monitoring & Employed Markov Chain with parameter adaptation  \cite{bib30} & Demonstrated the effectiveness of the model in early detection of health anomalies \\
Anomaly detection in smart city sensor networks & Integrated Markov Chain with machine learning techniques  \cite{bib31} & Improved detection accuracy compared to traditional anomaly detection methods \\
Study on adapting Markov Chain for changing operational requirements & Developed an adaptive Markov Chain model \cite{bib32} & Emphasized model adaptability in dynamic sensor networks \\
Real-time anomaly detection in diverse sensor types & Utilized a hybrid approach combining Markov Chain and neural networks \cite{bib33} & Achieved real-time anomaly detection with minimal false positives \\
Wireless sensor network security using Markov Chain & Investigated the vulnerability of WSNs to security threats \cite{bib34} & Proposed a novel Markov Chain-based security framework for WSNs \\
Anomaly detection in agricultural sensor networks & Applied Markov Chain to analyze crop health data \cite{bib35} & Improved crop monitoring accuracy by identifying anomalies in sensor readings \\
Markov Chain-based anomaly detection in vehicular networks & Developed a model to detect abnormal behavior in vehicle sensor data \cite{bib36} & Enhanced safety and security in smart transportation systems \\
\hline
\end{tabular}
\end{table}

As WSNs continue to grow in size and complexity, scalability becomes a paramount concern. The scalability of existing anomaly detection techniques is often limited due to the computational and resource constraints nature of the nodes. To address this gap, there is a need for anomaly detection methods that can scale with the size of the network and adapt to evolving sensor configurations. In addition, WSNs are inherently dynamic and the characteristics of the sensor data can change over time. Existing models may not be easily adaptable to these changes without manual intervention, making them less suitable for long-term deployment. Developing anomaly detection methods that can adapt to evolving network conditions is an unexplored avenue. Although Markov chain models offer promise in capturing temporal dependencies, there is a gap in exploring their practical implementation and performance within the unique context of WSNs. This includes not only the modeling aspects but also the practical challenges of parameter estimation and optimization for efficient anomaly detection.

By addressing these gaps in the existing research, our study aims to contribute to the development of anomaly detection techniques in WSNs that are scalable to growing networks, adaptable to changing network conditions, and effectively leverage the Markov Chain model's capabilities for improved anomaly detection accuracy and reliability.

\section{Methodology} \label{section-3}
In this section, the design of the proposed anomaly detection framework based on the first-order Markov chain model is discussed. The approach is designed to capture normal data dynamics in sensor readings and identify anomalies as deviations from learned transition patterns \cite{bib37,bib38}. The methodology consists of four major components, namely data pre-processing, state representation, Markov model construction, and anomaly detection.
\subsection{Data processing } The experiments were carried out using the publicly available Intel Berkeley Research Lab dataset, which contains time-stamped sensor readings from 54 sensor nodes deployed across in an indoor environment of Intel research lab in early 2004 using 54 Mica2Dot sensors. SNs are placed at different locations with varying exposures to light, airflow, and human interaction. Each node records temperature, humidity, light intensity and voltage at an interval of 31 seconds \cite{bib41}. We selected Intel lab Data Set due to its scale, heterogeneity, and temporal resolution in the real-world, making it suitable for performing anomaly detection algorithms.  Each record includes the following features: date, time, epoch, mote ID, temperature, humidity, light, and voltage over a period of more than one month. We focused on Mote ID 6 which is located near HVAC equipment because of its history of abnormal temperature fluctuations. The temperature sensor readings, as this is the most sensitive to environmental and device level anomalies such as overheating, sensor failure, sudden drops and more. The dataset further enables correlation-based anomaly validation by referencing humidity, light, and voltage. 

The purpose of selecting a single node is to evaluate how effectively the proposed framework identifies abnormal patterns and distinguishes between environmental variations and hardware-induced faults. To focus on temporal behavior, the dataset is first cleaned and filtered. Following steps were carried out in order to perform pre-processing the dataset.

\begin{itemize}
    \item Missing or corrupt entries are removed.
    \item Timestamps are constructed by combining the date and time fields into a single datetime index.
    \item Sensor readings are grouped per node, and temperature values are prioritized for anomaly detection due to their sensitivity to environmental changes.
    \item The data is resampled to hourly intervals using mean aggregation to smooth short-term fluctuations.
\end{itemize}
\subsection{State representation}
Markov chains require a finite set of discrete states. To convert the continuous temperature values into discrete states, we employ quantile-based binning, which divides the range of values into N bins (states) with approximately equal sample sizes \cite{bib39}. Each temperature reading is assigned to a corresponding state:
\begin{equation}
    S={S_1,S_2,...,S_N}
\end{equation}
This step transforms the continuous time series into a sequence of discrete observations suitable for Markov modeling. The notation used in the paper is discussed.
\begin{itemize}
    \item  $X_t$ represents the state of the sensor network at time $t$, where $X_t \in \{1, 2, \ldots, N\}$, with $N$ being the number of possible states.
    \item $Y_t$ represents the sensor reading at time t.
    \item $P(X_t = i)$ is the probability that the system is in state i at time t.
    \item $P(X_t = i, X_{(t-1)} = j)$ is the transition probability from state j to state i.
    \item $P(Y_t \mid X_t = i)$ is the Probability Distribution Function (PDF) of the sensor reading $Y_t$, since the system is in state i.
    \item $\theta_i$ is a parameter vector that characterizes the distribution of $Y_t$ when $X_t = i$.
\end{itemize}
\subsection{Markov Chain construction}
A first-order Markov chain assumes that the probability of transitioning to the next state depends only on the current state \cite{bib40}. Given a sequence of states $\{S_1,S_2,...,S_N\}$, the model constructs a transition probability matrix (TPM) P, where:
\begin{equation}
    P_{ij}=P_{(S_{t+1}=j  \mid  S_t=i)} 
\end{equation}

The matrix P is estimated from historical sequences by computing the relative frequency of transitions between state pairs. This captures the normal temporal behavior of the sensor under stable environmental conditions.

\subsection{Anomaly detection}
An anomaly is defined as a transition between states that occurs with low probability under the learned Markov model. Given the transition from state $s_i$ to $s_j$, the corresponding probability $P_{ij}$ is queried from the TPM. If $P_{ij}$ falls below a predefined threshold $\theta$, the transition is flagged as anomalous:
\begin{equation}
    \text{Anomaly if } P_{ij} < \theta
\end{equation}

The threshold $\theta$ is a tunable parameter and can be adjusted to balance sensitivity and specificity, depending on the requirements of the application. In this study, it is empirically selected based on validation performance.
The proposed model is given in algorithm \ref{algorithm1}.

\begin{algorithm}
\caption{Markov Chain Based Anomaly Detection} \label{algorithm1}
\begin{algorithmic}[1]

\Procedure{Build\_Transition\_Matrix}{state\_sequence, $k$}
    \State $C \gets$ zero\_matrix($k$, $k$) \Comment{Initialize transition count matrix}
    \For{$t = 0$ to $\text{len}(state\_sequence) - 2$}
        \State $i \gets state\_sequence[t]$
        \State $j \gets state\_sequence[t+1]$
        \State $C[i][j] \gets C[i][j] + 1$
    \EndFor
    \State $P \gets$ normalize\_rows($C$) \Comment{Convert counts to probabilities}
    \State \Return $P$
\EndProcedure

\vspace{0.5em}
\Procedure{Calculate\_Likelihood}{sequence, $P$}
    \State $likelihood \gets 1.0$
    \For{$t = 0$ to $\text{len}(sequence) - 2$}
        \State $i \gets sequence[t]$
        \State $j \gets sequence[t+1]$
        \State $likelihood \gets likelihood \times P[i][j]$
    \EndFor
    \State \Return $likelihood$
\EndProcedure

\vspace{0.5em}
\Procedure{Anomaly\_Detection}{state\_sequence, $P$, window\_size, $\varepsilon$}
    \State $T \gets \text{len}(state\_sequence)$
    \State $anomalies \gets$ empty\_list()
    \For{$t = 0$ to $T - window\_size$}
        \State $window \gets state\_sequence[t : t + window\_size]$
        \State $likelihood \gets$ \Call{Calculate\_Likelihood}{$window, P$}
        \If{$likelihood < \varepsilon$}
            \State $anomalies.append(\text{True})$
        \Else
            \State $anomalies.append(\text{False})$
        \EndIf
    \EndFor
    \State \Return $anomalies$
\EndProcedure

\end{algorithmic}
\end{algorithm}

The proposed model is evaluated on the Intel dataset using the following metrics.
\begin{itemize}
    \item Anomaly count compared to domain-based expectations.
    \item True vs. false detections, validated using visual inspection and known sensor failures.
    \item Comparison with baseline methods including:   \begin{itemize}
        \item Statistical Z-score method
        \item Autoencoder reconstruction error
        \item Hidden Markov Models (HMM)
    \end{itemize}
\end{itemize}

\begin{figure}
    \centering
    \includegraphics[width=0.6\linewidth]{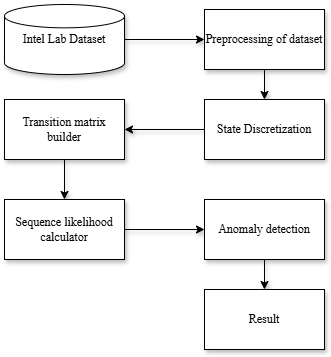}
    \caption{Flow of control in the proposed framework}
    \label{fig:fig2}
\end{figure}

The framework given in Figure \ref{fig:fig2} is implemented using Python with Pandas and NumPy for data processing and Matplotlib for visualization. The proposed approach is unsupervised, fast to train and interpretable, making it well suited for deployment on real-world WSNs with limited computational capabilities. 

\section{Results and discussion}
\begin{figure}
    \centering
    \includegraphics[width=0.8\linewidth]{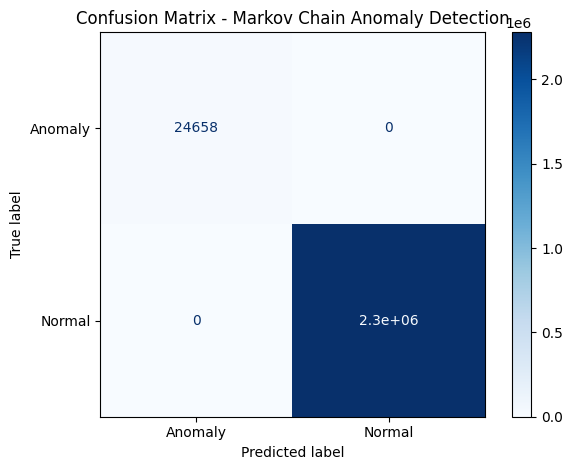}
    \caption{Confusion matrix}
    \label{fig:fig3}
\end{figure}

This section presents the experimental setup, implementation details, and evaluation of the proposed Markov chain-based anomaly detection model. To validate anomaly detection, a combination of manual labeling based on domain knowledge and historical patterns was used. Data points that reflected unrealistic environmental conditions or matched known faulty sensor behavior were labeled anomalies. No synthetic anomalies were introduced to preserve the integrity of the dataset. A confusion matrix was generated to evaluate model performance, showing high precision and a low false positive rate in Figure \ref{fig:fig3}. Known misbehaving node (e.g., Node 6) was used to qualitatively confirm true positives. The results are analyzed in terms of anomaly detection performance, scalability, and comparison with state-of-the-art techniques.
\subsection{ Experimental Setup}
Node Mote ID 6 from the dataset was selected due to its location, which is near to HVAC equipment. The node is selected for focused analysis because of its known history of abnormal temperature fluctuations. The temperature readings from this node were discretized into five quantile-based states (S1–S5), and the transition probability matrix  was derived from historical normal behavior. The quantile-based bins (0, 0.25, 0.5, 0.75, 1.0) are used to represent discrete states for the Markov chain. The detection threshold was set at $\theta$ = 0.05, and anomalies were flagged whenever an observed transition probability fell below this threshold. 

\subsection{Transition Probability Matrix (TPM) construction} The TPM given in table \ref{TPMtab1} was constructed for the selected node. The TPM describes how a system modeled by the proposed method is to move from one state to another over a single time step. Each element of the matrix represents the conditional probability that the process will move to state j given it is in state i. The row represents current state and the column to the future state. All probabilities are non-negative and each row sums to 1 because all possible next-state transitions exhaust the probability space. A global threshold $\theta$ was used to detect low-probability transitions. Low-frequency transitions (e.g., $\theta \leq$ 0.05) are considered potentially anomalous. 
\begin{table}[h]
\caption{State Transition Probability Matrix}\label{TPMtab1}%
\begin{tabular}{@{}llllll@{}}
\toprule
From / To & S1  & S2 & S3 & S4 & S5\\
\midrule
\textbf{S1} & 0.60 & 0.30 & 0.05 & 0.05 & 0.00 \\
\textbf{S2} & 0.10 & 0.70 & 0.15 & 0.05 & 0.00 \\
\textbf{S3} & 0.05 & 0.10 & 0.75 & 0.10 & 0.00 \\
\textbf{S4} & 0.00 & 0.00 & 0.20 & 0.70 & 0.10 \\
\textbf{S5} & 0.00 & 0.00 & 0.00 & 0.15 & 0.85 \\
\botrule
\end{tabular}
\end{table}

High transition probabilities in P11, P22 and P33 are 0.60, 0.70 and 0.75 respectively, indicate that once a system enters a state, it tends to remain there and represents system stability in states S1-S3. Moderate transition chances like P24=0.05, P24 =0.05 or P43=0.20 P43 =0.20 capture the system’s ability to move between adjacent states, representing gradual environmental or operational variability. The transition from S4 to S5 (0.10) and S5 to S4 (0.15) indicates a low but possible oscillation between these high states, suggesting transient anomalies before stabilization.

This TPM models the state evolution of temperature readings at the chosen SN. When transition probabilities fall below a defined threshold, the transition is considered unlikely, signaling a potential anomaly. High probabilities on the main diagonal represent normal behavior, while deviations from these steady transitions correspond to unexpected variations or sensor faults.

\subsection{Anomaly detection results} 
\begin{figure}
    \centering
    \includegraphics[width=0.9\linewidth]{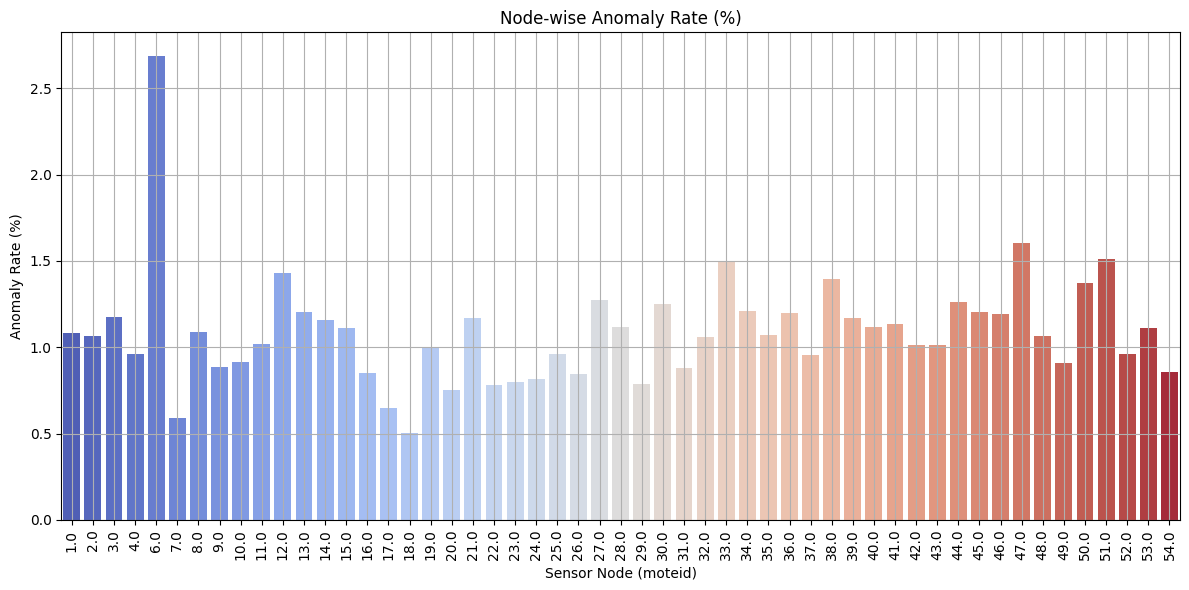}
    \caption{Node-wise anomaly rate (\%)}
    \label{fig:Fig4}
\end{figure}

\begin{figure}
    \centering
    \includegraphics[width=0.9\linewidth]{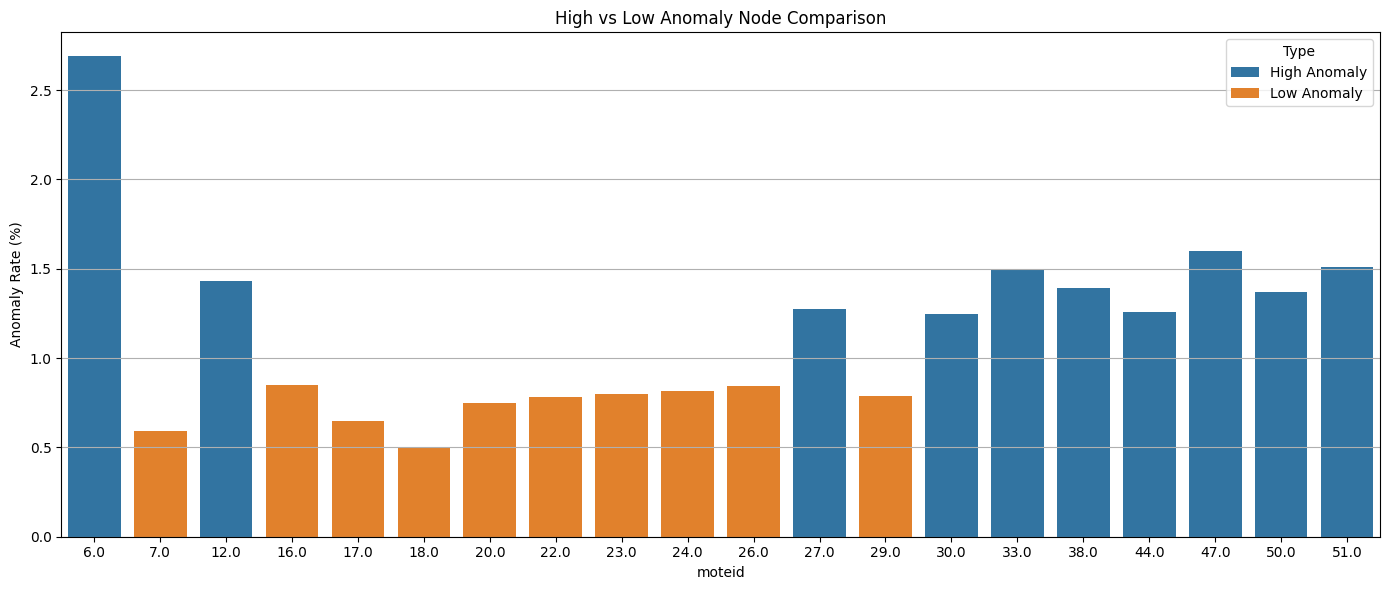}
    \caption{High and Low anomaly node comparison}
    \label{fig:Fig5}
\end{figure}

Anomalies were flagged when observed transitions had probabilities of deviating from the threshold $\theta$ = 0.05. Two distinct types of anomalies were detected namely thermal spikes and sudden drops. In first the abrupt increases in temperature while neighboring SN remained stable suggest that localized sensor malfunction drift. While in later short-duration dips in sub-ambient temperatures indicated transient packet losses. The detection results were validated by exploring correlated humidity and voltage readings. A decline in voltage synchronized with false temperature drops confirmed that those anomalies were hardware related rather then environmental. Figure \ref{fig:Fig4} shows the percentage of node-wise anomaly rates, while Figure \ref{fig:Fig5} shows ten high and low anomaly nodes.

\subsection{Comparison with other methods}
The proposed model was compared with several commonly used anomaly detection techniques in Table \ref{PCADMtab1}. The proposed Markov chain model demonstrate a balance between evaluation parameters and computational efficiency among the evaluated methods for anomaly detection in WSNs. In Figure-6, when compared to Z-score thresholding method which requires very low computation time however yields low precision and recall, the proposed Markov model achieves higher scores in state transitions. The HMMs performs better in precision, recall and F1-score but requires high computational resources which limit their applicability in real-time resources constrained WSN. Autoencoders offer the highest precision but requires very high computation resources and long training times. The proposed Markov chain model is interpretable, lightweight, and well-suited for distributed deployment on sensor nodes. The proposed model provides robust anomaly detection performance without the overhead of model training or extensive parameter tuning, making it a practical and scalable option for real-time anomaly detection for WSNs.
\begin{table}[h]
\caption{Performance Comparison of Anomaly Detection Methods}\label{PCADMtab1}%
\begin{tabular}{@{}lllll@{}}
\toprule
Method & Precision  & Recall & F1-Score & Computation Time\\
\midrule
Markov Chain (proposed) & 0.89 & 0.83 & 0.86 & Low \\ 
Z-score Thresholding & 0.73 & 0.65 & 0.69 & Very Low \\ 
Hidden Markov Model & 0.91 & 0.85 & 0.88 & High \\ 
Autoencoder & 0.94 & 0.82 & 0.87 & Very High \\
\botrule
\end{tabular}
\end{table}

The performance comparison reflects several benefits of the proposed model such as early detection of drifts and failure patterns. The state-transition representation provides an intuitive probabilistic explanation of anomalies which makes is interpretable. The proposed model is resource efficient as model execution and inference require negligible computational overhead, allowing on-node deployment even in low-power WSN environments. The proposed model is scalable because the TPM can be computed independently per node, facilitating distributed anomaly analysis without a central server.

\section{Conclusion and future work}
This study presents an effective and interpretable approach to anomaly detection in WSNs using a first-order Markov chain model. The indoor environmental monitoring case establishes that the proposed framework not only achieves robust detection accuracy but also enables explainable anomaly reasoning. Unlike deep neural models, the state-transition-based logic allows operators to trace the origin of irregularities, making it valuable for real-time fault isolation and network maintenance. This experiment validates the framework’s potential for smart building automation, industrial machine monitoring, and smart agriculture scenarios. By modeling the transitions between discretized temperature states, the proposed method identifies anomalies based on the probability of observed state sequences. Experiments with real-world data from the Intel Lab sensor deployment demonstrate that this method can reliably detect abnormal behavior while maintaining low computational complexity. Compared to traditional statistical methods and more complex machine learning models, the Markov chain model achieves a balanced trade-off between accuracy, transparency, and computational efficiency. The method is particularly well-suited to resource-constrained environments typical of WSNs, where explainability and real-time performance are crucial.

The model successfully captures temporal dependencies and detects both point and contextual anomalies without requiring labeled data or extensive training, high-lighting its applicability for real-time unsupervised anomaly detection across various domains. Although the current model looks promising, several directions for improvement and extension are identified. The proposed model has few limitations such as fixed threshold selection, single-feature focus, first-order Markov assumption, limited retrospective analysis, and no evaluation on generalization. In the study, the anomaly detection relies on a likelihood selection threshold which is manually selected and fixed.  Although the threshold is empirically tuned but may not generalize well across different datasets or network conditions. Use of adaptive threshold may improve the performance. Another limitation of the proposed model is the use of single feature while the model is capable of handling multivariate data. Use of multiple features will enhance the detection of contextual anomalies. Further more the model uses first-order dependencies which are suitable for capturing long term temporal dependencies that are critical for periodic anomalies detection.  In addition, the proposed model is applied in batch manner and is not useful for real-time performance evaluation and the model is tested only on the Intel lab dataset which limits is evaluation with different sensors, network density and sampling frequencies. 

The Markov Chain-based framework is a promising approach for lightweight, interpretable, and scalable anomaly detection in WSNs. It can adjust state boundaries based on real-time data distributions, potentially improving sensitivity to evolving environmental conditions. The framework can incorporate higher-order Markov Chains to capture more complex behaviors and enhance the detection of subtle anomalies. It can also integrate with spatial models, enabling the detection of regional anomalies or coordinated attacks. The model can be implemented on actual WSN nodes or edge devices to validate its performance under operational constraints. It can also serve as a lightweight first-stage filter in hybrid anomaly detection pipelines, flagging suspicious transitions for further analysis using more complex models. The framework can also be extended to multivariate sequences, improving accuracy and robustness in real environments.

\backmatter

\bmhead{Acknowledgement}
We acknowledge all helping brains in their direct and indirect support for this study.
\bmhead{Conflict of Interest}
The authors declare no conflict of interest.

\bmhead{Funding Information}

This research received no external funding.

\bmhead{Data Availability}

The original contributions are included in the article, inquiries can be directed to the corresponding author. 

\bmhead{Author Contribution Statement}
Conceptualization: RM, SKJ, NK; Methodology: RM, BB, SS; Analysis and Investigation: RM, SKJ, OFO, NK; Writing: RM, SKJ, OFO, BB, SS, NK; Visualization and Supervision: SKJ, SS, NK

\section*{Author Biography}
\begin{minipage}{0.3\textwidth}
  \includegraphics[width=\linewidth]{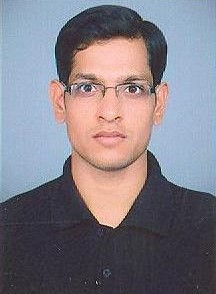} 
\end{minipage}%
\begin{minipage}{0.65\textwidth}
  \textbf{Rahul Mishra} 
  Mr. Rahul Mishra holds an M. Tech. in Computer Science and Engineering from Department of Computer Science and Engineering, MNNIT, Prayagraj, India. His passion for technology brings his into research and is currently pursuing Ph. D.  from the Department of Electronics and Communication, University of Allahabad, Prayagraj, India. His major area of interest includes Wireless Sensor Networks, Blockchain, and detecting and preventing handheld devices from Botnet. He has published numerous research papers and book chapters in journals of international repute. His work is driven to optimize the use of available resources to enhance the productivity. Through his research, he aims to develop cutting-edge solutions that address emerging cybersecurity challenges.
\end{minipage}

\begin{minipage}{0.3\textwidth}
  \includegraphics[width=\linewidth]{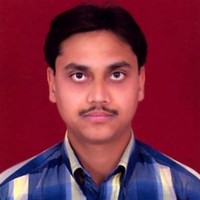} 
\end{minipage}%
\begin{minipage}{0.65\textwidth}
  \textbf{Sudhanshu Kumar Jha} 
  Dr. Jha is Assistant Professor in the Department of Electronics and Communication, University of Allahabad, India. He earned his Ph.D. from IIT (ISM) Dhanbad in 2011 and has over 13 years of teaching and research experience. He holds expertise in artificial intelligence, wireless sensor networks, blockchain-based systems, and emerging technologies in education. His research interests include machine learning, multimodal emotion intensity detection, secure communication in sensor networks, and the development of AI-driven tools for real-world applications. He has authored two textbooks, published extensively in reputed SCI journals and conferences, and holds nine patents. Dr. Jha is a recipient of several awards, and actively contributes to academic outreach programs. With a passion for integrating technology into teaching and learning, he works towards bridging the gap between theoretical research and practical innovation.
\end{minipage}

\begin{minipage}{0.3\textwidth}
  \includegraphics[width=\linewidth]{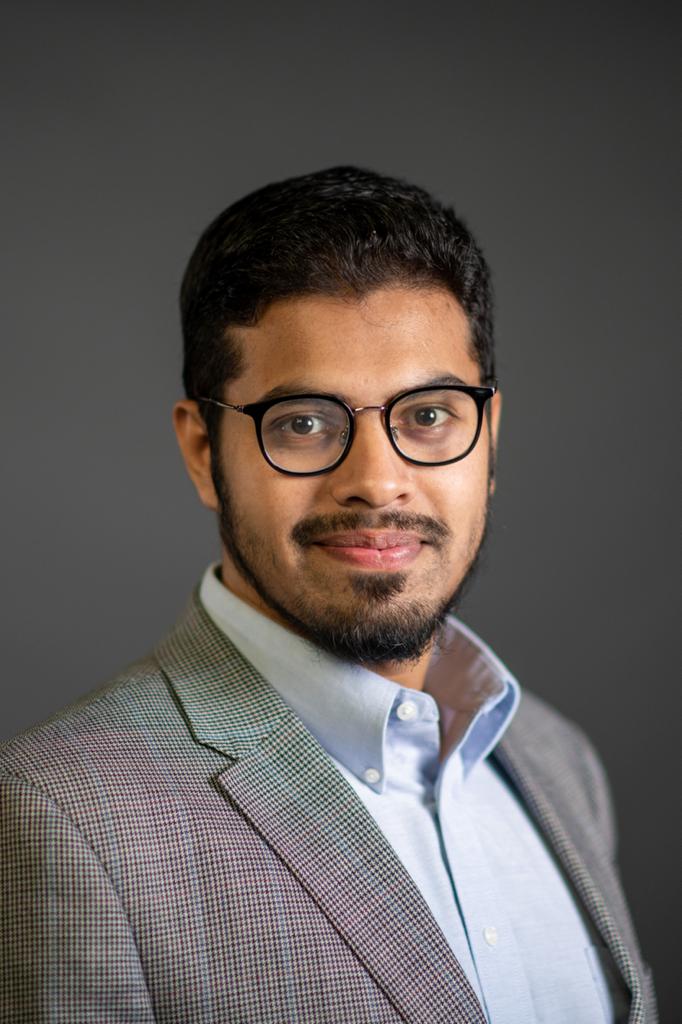} 
\end{minipage}%
\begin{minipage}{0.65\textwidth}
  \textbf{Omar Faruq Osama} 
  Omar Faruq Osama is a Data Scientist and Ph.D. researcher specializing in health systems engineering and clinical analytics. With over four years of experience in healthcare data management, predictive modeling, and digital transformation, Omar has delivered impactful solutions that improve patient care quality and operational efficiency. He has led research database development, integrated multimodal EMR datasets, and built AI-driven models supporting early clinical interventions and policy decisions. Omar is recognized for translating complex data into actionable insights while maintaining the highest standards of data governance, HIPAA/IRB compliance, and stakeholder collaboration. He is currently pursuing his Ph.D. in Industrial and Systems Engineering (Health Systems track) with a mission to advance data-driven healthcare.
\end{minipage}

\begin{minipage}{0.3\textwidth}
  \includegraphics[width=\linewidth]{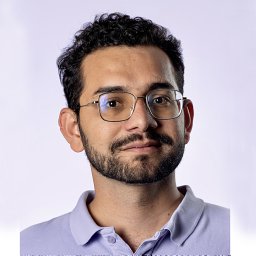} 
\end{minipage}%
\begin{minipage}{0.65\textwidth}
  \textbf{Bishnu Bhusal} 
  Bishnu Bhusal is a Ph.D. candidate in Computer Science at the University of Missouri–Columbia, where he also earned his Master’s degree with a focus on Cybersecurity and a minor in Statistics. His research interests include formal methods, privacy, cybersecurity, and machine learning, and his work has been published in leading venues such as CCS, OOPSLA, and GameSec.
\end{minipage}

\begin{minipage}{0.3\textwidth}
  \includegraphics[width=\linewidth]{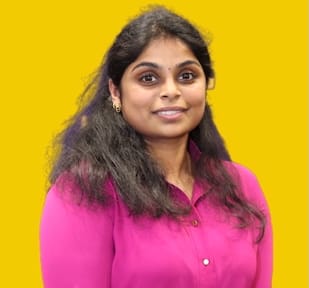} 
\end{minipage}%
\begin{minipage}{0.65\textwidth}
  \textbf{Sneha Sudhakaran} 
  Dr. Sneha Sudhakaran is a distinguished scholar and cybersecurity expert currently serving as an Assistant Professor in the Department of Computer Science at the Florida Institute of Technology. Dr. Sudhakaran’s research interests encompass various cybersecurity topics, including Android Security, Application Security, Host Security, Cyber Forensics, and Blockchain. As a cyber forensic researcher, she aims to create more profound knowledge for cyber forensic courses for students at the Florida Institute of Technology, which would help students get better jobs. Her certifications further validate her expertise as a Certified Ethical Hacker (CEH) and a Certified Hacking and Forensic Investigator (CHFI). Dr. Sudhakaran’s dedication to advancing cybersecurity education and practice is evident in her ongoing research. Her work not only contributes to academic knowledge but also provides practical solutions to real-world cybersecurity challenges, thereby fostering a more secure digital environment.
\end{minipage}

\begin{minipage}{0.3\textwidth}
  \includegraphics[width=\linewidth]{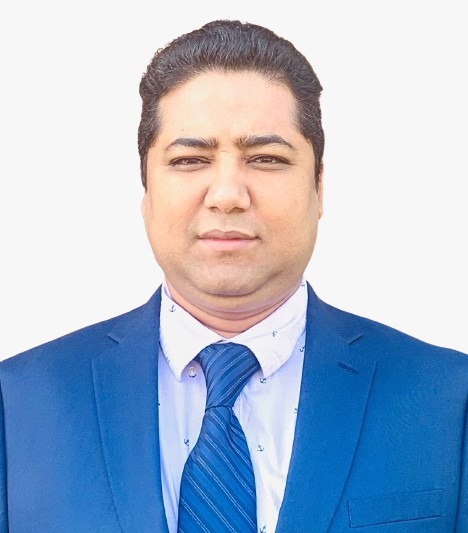} 
\end{minipage}%
\begin{minipage}{0.65\textwidth}
  \textbf{Naresh Kshetri} 
  Dr. Naresh Kshetri (BCA ’10, MCA ’14, MS ’17, PhD ’22) is currently a full-time Lecturer / Faculty (Cybersecurity) at Golisano College of Computing and Information Sciences (GCCIS), Rochester Institute of Technology (RIT). Dr. Kshetri completed his PhD (Computer Science with concentration in Cybersecurity) from University of Missouri – St. Louis, graduated with an MS (Cybersecurity) from Webster University, and also earned an MCA (Computer Applications) degree from University of Allahabad. With more than eleven years of experience and research interests in Cybersecurity, AI Security, Blockchain technology, he has published in various journals, conferences, and book chapters. His research is funded by the University of Missouri – St. Louis, the Lindenwood University, and the Emporia State University. Dr. Naresh has taught various Cybersecurity and CS courses at both the undergraduate and graduate levels. He has served as a reviewer (invited) for international journals (IEEE TNSM, IEEE OJVT, Wiley ETT, PeerJ CS and so on), conferences (IEEE ICICT, IEEE AIBThings, IEEE ETNCC, ACM SIGCITE, AOM, and so on), books (CRC Press, Elsevier, IGI Global), and on the Technical Program Committee (TPC including Session Chairs and International Advisory Committees) member of prestigious conferences. He is a Senior Member of IEEE (M ’19, SM ’25) and a Professional Member of ACM (M ’25). For more details about Dr. Naresh Kshetri, please visit his website: https://sites.google.com/view/nareshkshetri  
\end{minipage}
\end{document}